\documentclass[twocolumn,apj]{emulateapj}

\usepackage{graphicx}
\usepackage{natbib}
\usepackage{amsmath}
\usepackage{amssymb}
\usepackage{fancyhdr}
\usepackage{latexsym}
\usepackage{color}
\usepackage{txfonts}
\usepackage{textalpha}

\begin{document}

\title{A parameter study for modeling \ion{Mg}{2} h and k emission during solar flares}

\author{Fatima Rubio da Costa\altaffilmark{1}, Lucia Kleint\altaffilmark{2}}
\altaffiltext{1}{Department of Physics, Stanford University, Stanford, CA 94305, USA; Email: frubio@stanford.edu}
\altaffiltext{2}{University of Applied Sciences and Arts Northwestern Switzerland, 5210 Windisch, Switzerland}

\begin{abstract}
{Solar flares show highly unusual spectra, in which the thermodynamic conditions of the solar atmosphere are encoded. Current models are unable to fully reproduce the spectroscopic flare observations, especially the single-peaked spectral profiles of the \ion{Mg}{2}~h~and~k~lines. We aim at understanding the formation of the chromospheric and optically thick \ion{Mg}{2}~h~and~k~lines in flares through radiative transfer calculations.}
{We take a flare atmosphere obtained from a simulation with the radiative hydrodynamic code RADYN as input for a radiative transfer modeling with the RH code. By iteratively changing this model atmosphere and varying thermodynamic parameters, such as temperature, electron density, and velocities, we study their effects on the emergent intensity spectra.}
{We can reproduce the typical single-peaked \ion{Mg}{2}~h~and~k flare spectral shape and their approximate intensity ratios to the subordinate \ion{Mg}{2} lines by either increasing densities, temperatures or velocities at the line core formation height range.}
{Additionally, by combining unresolved up- and downflows up to $\sim$250~km~s$^{-1}$ within one resolution element, we also reproduce the widely broadened line wings. While we cannot unambiguously determine which mechanism dominates in flares, future modeling efforts should investigate unresolved components, additional heat dissipation, larger velocities, and higher densities, and combine the analysis of multiple spectral lines.}
\end{abstract}

\keywords{Sun: flares; chromosphere --- line: profiles --- radiative transfer --- hydrodynamics (HD)}

   \section{Introduction}\label{Sect:intro}
Although solar flares have been investigated for decades, there are still open questions, such as how the released energy is transported throughout the atmosphere and deposited in the lower atmospheric layers. By studying chromospheric spectral lines, we probe the underlying physical processes and the response of the lower atmosphere to flare heating. NASA's most recent solar mission, the {\it Interface Region Imaging Spectrograph} \citep[{\it IRIS};][]{2014SoPh..289.2733D} showed puzzling spectra of the near-UV (NUV) \ion{Mg}{2}~h~and~k resonance lines in flares, which could not be explained by any modeling efforts so far.

The \ion{Mg}{2}~h~and~k lines are an important contributor to the UV emission during flares \citep{1984SoPh...90...63L}. Their unexplained characteristics in flares are: 1)~a lack of their central self-reversal, which is unusual for lines dominated by scattering; 2)~very broad line wings (line center~$\pm$~1.5~\AA); 3)~the presence of the subordinate \ion{Mg}{2} lines (3p-3d transition) in emission, which in the quiet Sun have been reported to be sensitive to heating in the low chromosphere \citep{2015ApJ...806...14P}, but may form differently in flares; and 4)~redshifts often occurring due to strong downward velocities \citep{2015A&A...582A..50K, 2015SoPh..290.3525L, 2015ApJ...807L..22G}.

The so far only direct comparison of observed flare \ion{Mg}{2} spectra and hydrodynamic simulations \citep{2016ApJ...827...38R} showed that the peculiar shape of flare spectra cannot be reproduced yet and that simulations always obtain profiles with a central reversal, more similar to quiet Sun than to flares. There have been several attempts to understand the behavior of the \ion{Mg}{2} spectra during solar flares \citep{1980ApJ...242..336M, 1983A&A...125..241L, 2016ApJ...827..101K, 2016ApJ...827...38R, 2015SoPh..290.3525L, 2016ApJ...830L..30D, 2017ApJ...836...12K, 2017arXiv170104213R}, but no simulation to date has reproduced any of the observed profiles and it is unclear, which physical mechanisms are responsible for the observed line profiles. Our goal for this paper is to carry out a parameter study to investigate the origin of the \ion{Mg}{2} spectral shapes during flares and to understand the atmospheric conditions needed to reproduce the observed flaring UV spectra. \citet{2015ApJ...809L..30C} performed a similar parameter study with the RH code, modifying the thermodynamic parameters in their model atmosphere, but enforcing hydrostatic equilibrium, to study the \ion{Mg}{2}~k profiles in plages with the objective of matching the observations. Here, we investigate even unlikely possibilities of variations, such as velocities or microturbulences significantly higher than those reported in observations, but we cannot provide an unambiguous solution to the conditions in a flaring atmosphere, which would require comparisons with multiple spectral lines and which will be our next step.

There are currently three codes commonly used for flaring hydrodynamic simulations: RADYN \citep{1997ApJ...481..500C}, FLARIX \citep{2009A&A...499..923K}, and HYDRAD \citep{2013ApJ...770...12B}. The HYDRAD code does not account for optically-thick radiative losses, an important energy term in the chromosphere. Therefore, it is not a suitable tool to study the response of the lower atmosphere to flare heating and in particular the \ion{Mg}{2} emission in the UV. The FLARIX code considers radiative transfer for hydrogen, calcium and more recently magnesium, but currently it does not consider the radiative losses of the helium atom, which may be up to 10\% of the total energy radiative loss \citep{1989ApJ...341.1067M, 2009ApJ...702.1553L}, affecting the modeled chromospheric emission. In contrast, RADYN solves the hydrodynamic equations together with the detailed radiative transfer for the atoms dominating the radiative losses in the chromosphere; i.e. hydrogen, helium, calcium and magnesium. Here we therefore use a model atmosphere from a RADYN simulation to start with the most realistic flare atmosphere and vary its thermodynamic parameters to investigate their influence on the \ion{Mg}{2} lines. The non-LTE radiative transfer code RH \citep{2001ApJ...557..389U} is used to calculate the \ion{Mg}{2} lines from the modified atmospheres. The advantage of RH is the proper treatment of the effects of angle-dependent partial frequency redistribution (PRD), improving the assumption of complete frequency redistribution (CRD) used by RADYN, which has previously been demonstrated to play an important role in the formation of the \ion{Mg}{2}~h~and~k line profiles \citep[see e.g.][]{2013ApJ...772...89L, 2013ApJ...772...90L}. Additionally, RH accounts for possible frequency overlap from bound-bound transitions \citep{1991A&A...245..171R, 1992A&A...262..209R}, whereas RADYN does not.

\begin{figure}[!tb]
\centering
  \includegraphics[width=.5\textwidth]{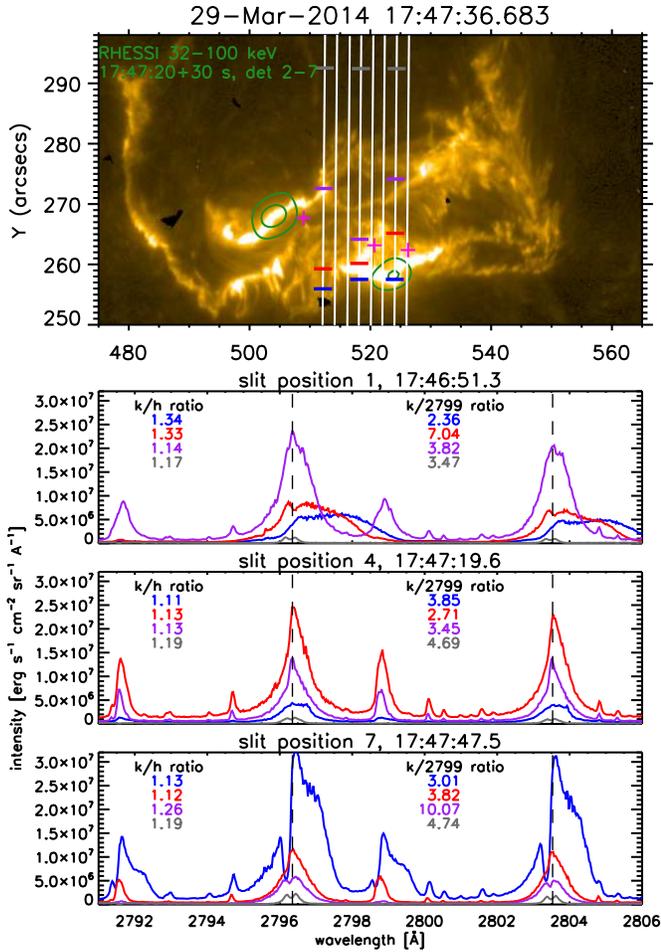}
   \caption{Top: {\it IRIS}~SJI~1400 image of the X1 flare on 2014 March 29 with the spectrograph slit during the 8 raster positions drawn as vertical lines. The green contours show RHESSI HXR emission and the magenta crosses denote the positions where H$\alpha$ and Ca 8542 \AA\ spectra were compared in \citet{2016ApJ...827...38R}. The bottom panels show selected \ion{Mg}{2} flare spectra. The colors indicate the location on the slits and correspond to the small horizontal lines drawn in the SJI image. Typical flare profiles have single peaks, but may also have broad red wings. The barely visible gray profiles show typical quiet Sun profiles. }
  \label{examplemg}
\end{figure}

The paper is organized as follows: a description of the motivation of our Mg line spectra study during solar flares is shown in Section~\ref{Sect:motivation}; a summary of the RADYN and RH codes is presented in Sections~\ref{sect:radyn} and~\ref{sect:rh}, followed by the study of the behavior of the \ion{Mg}{2}~h~and~k line profiles to different atmospheric conditions in Section~\ref{Sect:synthetic_profs}; a more detailed analysis on the line formation is described in Section~\ref{Sect:contr_fct}, and a summary and discussion of the results in Section~\ref{sect:conclusions}.

  \section{Motivation}\label{Sect:motivation}
The observed shapes of the \ion{Mg}{2} lines are very diverse and variations exist on pixel-to-pixel scales. Some examples from the X1 flare on March 29, 2014 \citep{2015ApJ...806....9K} are shown in Figure~\ref{examplemg}. Flare profiles range from single peaks with broad wings (e.g.\,middle panel red and purple spectra), to small central reversals (bottom panel purple), to strongly asymmetric red components (e.g. top panel red and blue, bottom panel blue). The locations of the profiles are indicated with color-coded horizontal lines in the image on top. Grey spectra show the quiet Sun for a comparison. {\it RHESSI} HXR contours from 32-100 keV are also drawn. Because the non-thermal counts were comparably low at this time, we needed to integrate for 30 s (17:47:20-17:47:50 UT), using detectors 2 to 7. The {\it RHESSI} contours are therefore not strictly simultaneous to the {\it IRIS} data, but the sources did not move significantly at this time \citep{2015ApJ...813..113B}. The blue profile of slit position 7 corresponds to a time just when accelerated electrons were detected at the same location. This spectral profile shows a reversal, though shifted to the blue, and additionally a downflow, and its intensity increases significantly. A few seconds later, such profiles turn into the broad single-peak profiles. For example, the HXR source passed the red and purple profiles of slit 4 about 30 to 90 seconds earlier.

The line ratios given in the images are calculated as integrals from 2795--2798 \AA\ (k line), 2802--2805 \AA\ (h line), and 2798--2800 \AA\ (subordinate line at 2799 \AA).

The magenta crosses are for reference only and indicate locations where the H$\alpha$ and \ion{Ca}{2} 8542 \AA\ were compared in \citet{2016ApJ...827...38R}. These spectral lines matched relatively well to the model atmosphere that is used for the parameter study in this paper.

Single-peaked profiles seem universal for all analyzed flares and are also commonly observed in sunspots, though in sunspot they appear without any broad wings and with only tiny, if any, emission from the subordinate lines. So our goal is to simulate single-peak broad \ion{Mg}{2}~h~and~k profiles (e.g. red or purple profile of slit position 4) and simultaneously produce emission in the subordinate 3p-3d \ion{Mg}{2} lines at the correct intensity ratios, to understand the characteristics of the atmospheric parameters necessary to reproduce the observed typical line profiles during solar flares.

   \subsection{Formation of the \ion{Mg}{2}~h~and~k Spectral Lines}{\label{Sect:formation_lines}}
The \ion{Mg}{2}~h~and~k lines are resonance lines formed under conditions of non-LTE. Considering that the radiative damping is the largest contribution to the total damping in the chromosphere (at $z>500$~km), that Van der Waals broadening is the dominant contribution in the photosphere and that quadratic Stark broadening is only of minor importance, the radiative damping component seems to be the main contribution to the formation of the \ion{Mg}{2}~h~and~k line core, while the Van der Waals broadening seems to be the main contribution to the broad line wings \citep{2013ApJ...772...89L}. 3D effects have been reported to be very important, especially in the line core. Unfortunately, there are no simulations yet that are able to calculate \ion{Mg}{2} realistically in 3D.

The 3p-3d level subordinate lines usually appear in absorption but show a significant enhancement during solar flares \citep[e.g.][]{2015A&A...582A..50K}. Their emission was analyzed by \citet{2015ApJ...806...14P} who concluded them to be optically thick and forming in the lower chromosphere in the quiet Sun. They found them to turn into emission only rarely, especially during their quiet Sun simulations, and concluded that a large temperature gradient ($\ge~1500$~K) must be present in the lower chromosphere to cause the subordinate lines to turn into emission.

  \section{Modeling the line profiles}\label{sect:radyn}
We start with a modeled flare atmosphere from a RADYN simulation and modify its parameters to investigate their influence on the \ion{Mg}{2} lines, using the non-LTE radiative transfer code RH to properly treat the PRD effects. In the following sections~\ref{sect:radyn} and \ref{sect:rh} we will briefly introduce each code individually.

  \subsection{RADYN Code}\label{sect:radyn}
We used the RADYN code of \citet{1997ApJ...481..500C}, including the modifications of \citet{1999ApJ...521..906A} and \citet{2005ApJ...630..573A}, to simulate the radiative-hydrodynamic response of the lower atmosphere to energy deposition by non-thermal electrons in a flare loop. We use the atmosphere from the run described in \citet{2016ApJ...827...38R} at 17:45:31~UT and manually modify its temperature, density, non-thermal broadening or plasma velocity with the RH code (see Section~\ref{sect:rh}). Additionally, we used RADYN simulations with electron beam fluxes from 10$^9$ to 10$^{12}$ erg s$^{-1}$ cm$^{-2}$ to investigate the dependence of the spectral line shape on beam flux. We also compared the spectra resulting from the RADYN simulations described in \citet{2015ApJ...813..133R}, including particle transport and stochastic acceleration with the ones assuming an ad-hoc single power-law, derived from fitting the non-thermal component of the hard X-ray spectra to a single power-law and applying the collisional thick-target modeling \citep{1971SoPh...18..489B, 1978ApJ...224..241E}.

Considering that the \ion{Mg}{2}~h~and~k lines are strongly affected by the effect of PRD, their simulations with RADYN, which only assume CRD, are not sufficiently realistic. We therefore recalculate the \ion{Mg}{2} lines with the RH code (see Section~\ref{sect:rh}) including PRD effects. As an input to RH we use snapshots of the atmosphere calculated with RADYN with manually modified parameters.

  \subsection{RH Code}\label{sect:rh}
We performed non-LTE radiative transfer computations with a modified version of the RH code \citep{2001ApJ...557..389U, 2015A&A...574A...3P}, which includes the heating due to the injection of electrons in the atmosphere. The modification (performed by J. Allred; private communication) includes the computation of the collisional ionization rate from non-thermal electrons, following \citet{1983ApJ...272..739R}, given the electron heating rate estimated from the RADYN code. RH can treat the effects of angle-dependent partial frequency redistribution using the fast approximation by \citet{2012A&A...543A.109L}. We use a 10 level plus continuum model atom for \ion{Mg}{2} as described in \citet{2013ApJ...772...89L}.

One of the differences between RH and RADYN is that RADYN uses the Uppsala opacity package \citep{Gustafsson}, while RH calculates the opacity of the transitions in the indicated atoms. Additionally, by using snapshots of the atmospheric model from RADYN as input into the RH code, we recalculate the ionization populations using statistical equilibrium, instead of using the non-equilibrium ionization already estimated by RADYN. This changes the plasma density and, as a consequence, shifts the height scale in comparison with the atmosphere produced by RADYN, leading to differences in the formation height in units of length. Although statistical equilibrium can be a poor assumption at the beginning of the impulsive phase of the simulation, the effect should be smaller at later times. Unfortunately there is currently no option to avoid statistical equilibrium in RH. We would like to emphasize that the goal is to study the behavior of the \ion{Mg}{2} line for different atmospheric parameter changes. Even though we use the ionization population based on the statistical equilibrium assumption for excitation processes, we still keep the non-equilibrium electron density estimated from RADYN, rather than recalculating the equilibrium electron density, since it is the one resulting from the hydrodynamic equations. Therefore, we believe it to be more realistic and the electron density population is far from equilibrium in a flaring atmosphere. By not solving for hydrostatic equilibrium, we have not enforced the charge conservation in RH. We find that the velocity variations in the atmosphere do not displace the charge balance of positive vs. negative charges significantly ($<$ 4\%). The temperature and density variations affect the charge conversation more with the exact numbers listed in the respective sections below. 

These points may make a difference when comparing the line profiles with the observations; for instance, the continuum emission at 2992~\AA\ resulting from RH is less than 20\% lower than the one estimated from RADYN. Another possibility that can explain the lower continuum emission estimated with RH, in comparison with RADYN, can be the lower opacity in the wings of the \ion{Mg}{2}~h and k lines. \citet{2017ApJ...836...12K} have included the \ion{Mg}{2}~h and k wing opacity in the calculations of the excess continuum intensity finding that the opacity at these wavelengths is important for an accurate treatment of the \ion{Mg}{2} wing emission as well as the continuum in the upper photosphere. An underestimation of the opacities can result in a lower continuum emission, by a factor of 15–-30\%.

Each line profile is calculated with up to 100 frequency points across the transition using a Voigt profile and a line broadening due to the Stark effect. Throughout the paper we use $\mu$=0.77 to be consistent with the observations in Figure~\ref{examplemg}.

\begin{figure*}[!tbh]
\centering
  \includegraphics[width=\textwidth]{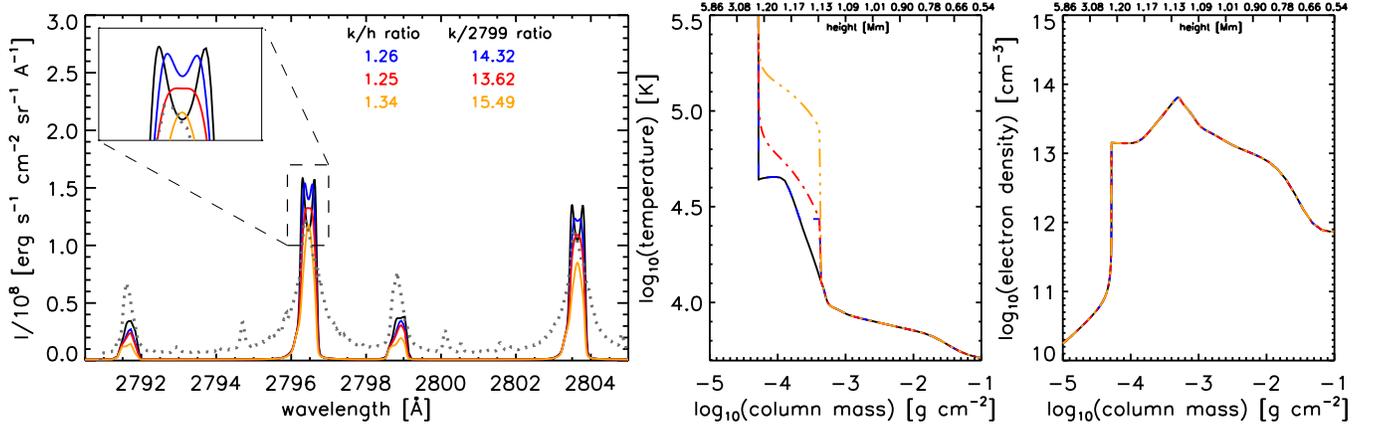}
   \caption{Variation of the temperature structure just below the transition region (middle panel) while keeping the electron density from RADYN (right panel) may lead to \ion{Mg}{2} lines with single peaks (e.g. yellow, left panel), similar to those observed in flares (example flare spectrum multiplied by 5 plotted as dotted gray line). The ratio between h and k lines to the subordinate lines ($\approx$ 15) is higher here than in the observations ($\approx$ 4). Note that the height scale is given in Mm is not linear in the plots because $\log_{10}$(column mass) is linear.}
  \label{vart}
\end{figure*}

\begin{figure*}[!htb]
\centering
  \includegraphics[width=\textwidth]{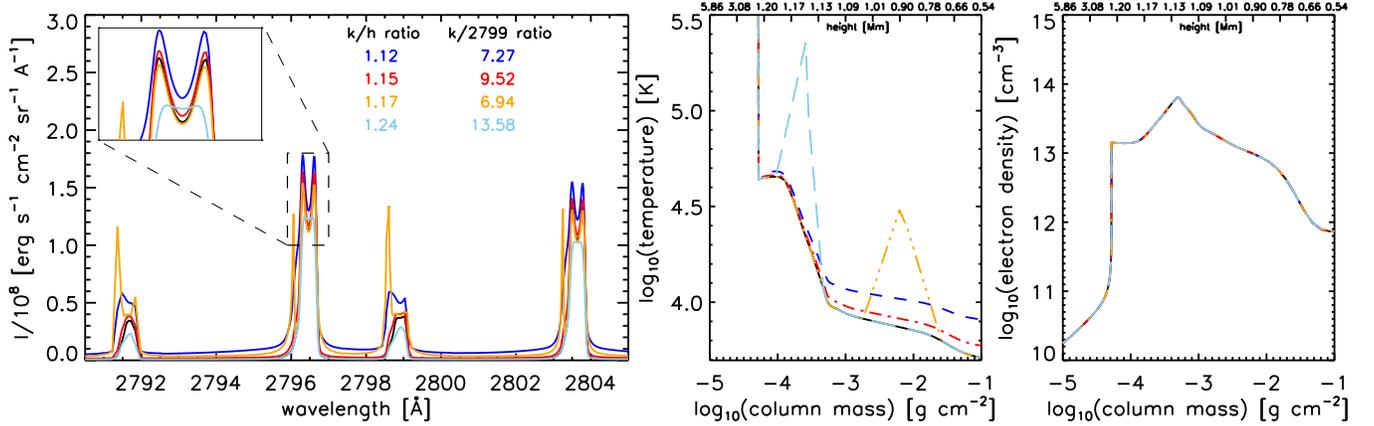}
   \caption{Increase of the chromospheric temperatures up to 3000~K (red, blue) slightly broaden the line wings, but not enough compared to the observations. Temperature spikes in the middle chromosphere lead to irregular unrealistic profiles (yellow). Temperature spikes in the upper chromosphere lead to single peak profiles. The temperature behavior (e.g. drop or increase) above the line core formation height does not influence the lines (light blue in comparison with Figure~\ref{vart}).}
  \label{truns}
\end{figure*}
  \section{Synthetic line profiles}\label{Sect:synthetic_profs}
In this section we modify one thermodynamic parameter at a time. While this probably is not representative of real solar changes during flares, it allows us to investigate the dependence of the \ion{Mg}{2} spectra on different parameters. We ran variations of several hundred atmospheres and show the main results in the following sections.

  \subsection{Varying temperature}\label{Sect:vart}
We varied the temperature structure, including {(1)}~temperature steps at different heights below the transition region, {(2)}~higher chromospheric temperatures, and {(3)}~localized heating at certain heights. The electron density and plasma velocities were kept constant for all runs. Figure~\ref{vart} shows some representative runs in which the temperature increases just below the transition region, where the core of the \ion{Mg}{2}~k and h lines are formed. A strong temperature increase in this region leads to single-peaked profiles whose intensity is lower than the original atmosphere. However, the intensities of the subordinate lines are too low compared to the h and k lines. The dotted gray line corresponds to the red spectrum at slit 4 in Figure~\ref{examplemg} with its intensity multiplied by 5. The observed line wings are clearly much broader.

We also increased the temperature of the chromosphere (red and blue lines in Figure~\ref{truns}) or introduced strong heating at limited heights (yellow and light blue lines in Figure~\ref{truns}). Even when unrealistically increasing the chromospheric temperature by 3000~K (blue), the line wings are only slightly broadened, far less than the observed broadening. Temperature spikes in the middle chromosphere (yellow) lead to sharp spikes in the line wings, which to our knowledge have not been found in observations. Temperature spikes in the upper chromosphere (light blue) lead to a similar behavior as shown in Figure~\ref{vart}. 

As long as there is a steep temperature increase at the formation height of the line core, the profile turns into single peak and the line is insensitive to the temperature behavior above these heights.

Because the charge conservation has not been enforced, the charge balance is displaced. Considering the atoms of hydrogen, helium, and magnesium, the maximum ratio between positive and negative charges is 1.16 at 10000~K and 1.01 at 50000~K in the models of Figure~\ref{vart}.

\begin{figure*}[!hbt]
\centering
  \includegraphics[width=\textwidth]{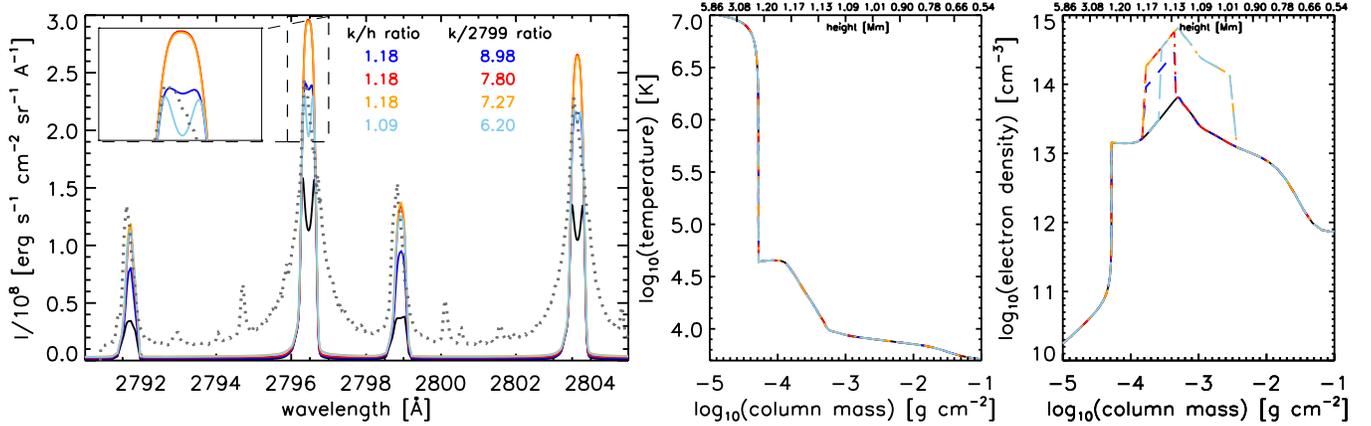}
   \caption{Variation of electron density. To obtain a single peak, a high density (factor of 10 increase) at the formation height of the line core (here -3.8 $\le$ log$_{10}$(column mass) $\le$ -3.6~g~cm$^{-2}$) is required as illustrated by the red and yellow examples. The intensity is too high compared to the observations, but the k/h ratio is similar and the k/2799 ratio is within a factor of 2. The dotted gray line is an example flare spectrum.}
  \label{vard}
\end{figure*}

\begin{figure*}[!htb]
\centering
  \includegraphics[width=\textwidth]{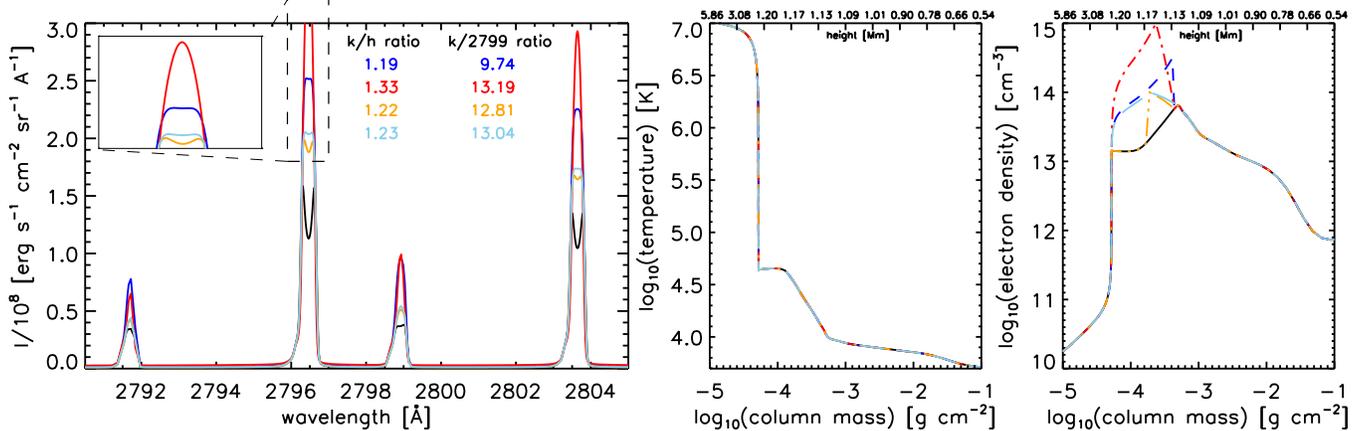}
   \caption{Linear increases of the density closer to the transition region than in Figure~\ref{vard}. A too high increase in density (red) not only leads to too high intensities, but also worsens the ratio of the k line vs. the subordinate lines.}
  \label{lind}
\end{figure*}

   \subsection{Varying electron density}\label{Sect:vard}
In this set of models, we kept the temperature constant and increased the electron density at different heights. Figure~\ref{vard} shows that heights just below the transition region have no effect on the shape of the line core (yellow and red lines). The comparison of yellow and light blue lines shows that the density between log(column mass) between $-3.6$ and $-3.8$~g~cm$^{-2}$, which corresponds to 1.15-1.17 Mm, is responsible for the single peak of the h and k lines. By increasing the density at least by a factor of 10, the reversed profile turns into a single peak (black vs. blue vs. red profiles). The k/h ratio is similar to the observations, the k/2799 ratio differs by a factor of 2, however the intensity is significantly too high in the models. Therefore the observed spectrum (dotted gray) was multiplied by a factor of 10.

Figure~\ref{lind} shows the effect of more smoothly increasing and decreasing densities. Similarly to the previous figure, only densities at the formation height of the line core matter for obtaining single-peaked profiles.

\citet{2017ApJ...836...12K} have recently studied the continuum emission together with the asymmetries observed by {\it IRIS} in the \ion{Fe}{2}~2832.39~\AA\ line profile during the impulsive phase of an X1.0 class flare. They suggest a model producing two flaring regions (a condensation and stationary layers) in the lower atmosphere. The condensation, due to the high densities in the chromosphere, compresses the chromosphere and therefore enhances the continuum emission; while the stationary flare layers in the chromosphere explain the bright redshifted \ion{Fe}{2} emission component, reported to be observed during solar flares. This plausible scenario with two flaring regions in the chromosphere, could explain the sudden density increase within a narrow region.

Another possible physical mechanism that can explain the increase of the electron density without elevating the temperature is the non-thermal ionization resulting from a beam of non-thermal electrons, creating a compressed region in the chromosphere for short timescales. The lack of charge conservation leads to unrealistic values (increase of negative charges by an order of magnitude at T=10000~K) in the yellow and light blue examples in Figure~\ref{vard}, because we modified the density starting low in the atmosphere. For all other atmospheres in Figure~\ref{vard} and \ref{lind}, the ratios are maximally 1.19 and 1.01 at T=10000~K and 50000~K, respectively.

\begin{figure*}[!htb]
\centering
  \includegraphics[width=\textwidth]{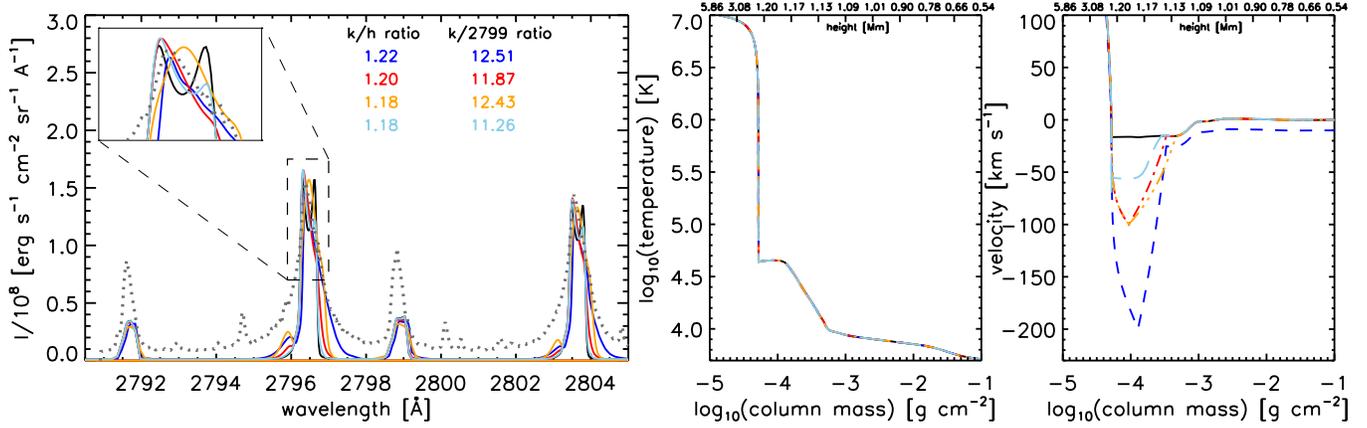}
   \caption{Variations of the Doppler velocities. Negative values denote redshifts. For the light blue example we approximated the velocity profile observed by \citet{2009ApJ...699..968M}. While single-peak profiles can be reproduced, the resulting spectral lines are not symmetric, contrary to the observations (dotted gray, multiplied by a factor of 6), and may include small extra components in the blue side of the h and k lines.}
  \label{varv}
\end{figure*}

   \subsection{Varying Doppler velocity}\label{Sect:varv}
The Doppler velocities may change strongly during flares, generally from strong upflows at coronal temperatures to downflows of few tens of km s$^{-1}$ in the chromosphere \citep[e.g.][]{2009ApJ...699..968M}. For these tests, we vary the magnitude of the downflow at and near the line core formation height. Figure~\ref{varv} shows the approximate Doppler velocities from \citet{2009ApJ...699..968M} in light blue (negative velocities represent downflows), which lead to an asymmetric profile with a stronger blue peak. It is possible to obtain a single-peak profile with its center at the line core just by varying velocities (e.g. yellow). But such variations do not lead to a symmetric profile.

\begin{figure*}[!htb]
\centering
  \includegraphics[width=\textwidth]{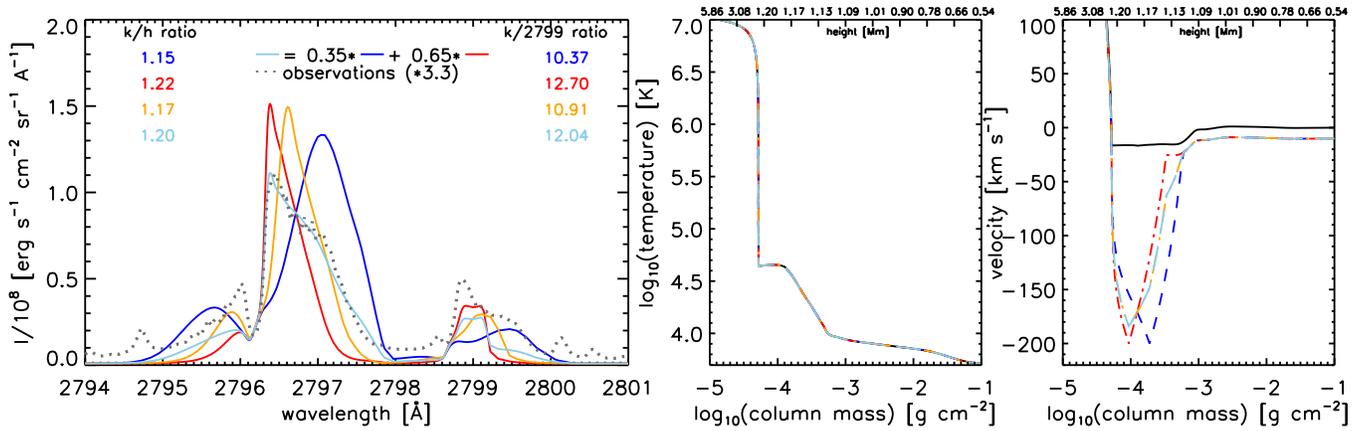}
   \caption{Variation of the Doppler velocities to match the blue profile from Figure~\ref{examplemg}. The red and dark blue profiles were combined to the light blue profile, which agrees well with the {\it IRIS} observations from near the HXR footpoint. The peak, the extended red wing, and the location of the blue peak at 2796~\AA\ can be reproduced, but the subordinate lines only roughly agree in shape, not intensity. The yellow profile shows the RH simulation using the combined velocity profile, indicating that a single component cannot match the {\it IRIS} observations and that there likely are unresolved downflows in one pixel.}
  \label{combine}
\end{figure*}

\begin{figure*}[!htb]
\centering
  \includegraphics[width=\textwidth]{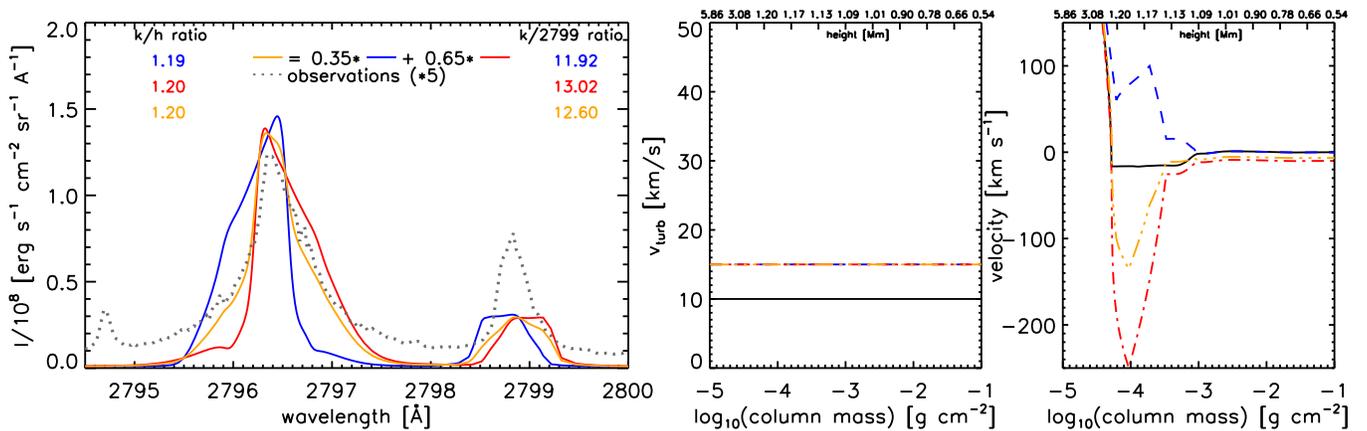}
   \caption{Variation of the Doppler up- and downflowing velocities to match the red profile from Figure~\ref{examplemg}, slit position 4. The red and dark blue profiles were combined to the yellow profile, which agrees well with the {\it IRIS} observations from the ribbon. The broad line wings, the shift of the line and the narrow peak, can be reproduced, but the modeled subordinate lines have lower intensities and broader profiles. Considering that such symmetric profiles cannot be obtained with any 1D velocity structure, this may indicate that unresolved up- and downflows exist in the same pixel.}
  \label{combine2}
\end{figure*}

All single-peak profiles plotted in Figure~\ref{varv} (yellow, blue, red) show an extra enhancement at the blue side of the line, at around 2795.9~\AA, similarly to the blue wing enhancement observed by {\it IRIS} in the bottom panel of Figure~\ref{examplemg} (blue profile). Such {\it IRIS} profiles seem to only occur during a few seconds and exactly at locations where HXR footpoints are observed. Figure~\ref{combine} shows the ``HXR-footpoint'' {\it IRIS} profile in dotted gray. We noticed that by starting the velocity variation lower in the chromosphere the emission becomes red-shifted (dark blue vs. red in Figure~\ref{combine}), but the location of the dip at $\sim$2796~\AA\ remains. We combined the red and blue profiles (35\% blue + 65\% red) into a new profile (light blue), simulating what would be observed if these two components existed within one pixel. The light blue profile qualitatively matches the ``HXR-footpoint'' {\it IRIS} profile very well, both in the shape and extent of the red wing and in the location of the dip. We re-ran RH with the velocities derived from this combination (yellow), but the resulting profile shows that a single component cannot explain the observations. While the non-perfect fit of the emission feature at $\sim$2796~\AA\ and of the subordinate lines indicates that the middle-lower chromosphere is not fully reproduced yet, our results indicate that ``HXR-footpoint'' {\it IRIS} profiles are due to unresolved strong downflows.

\begin{figure*}[!htb]
\centering
  \includegraphics[width=\textwidth]{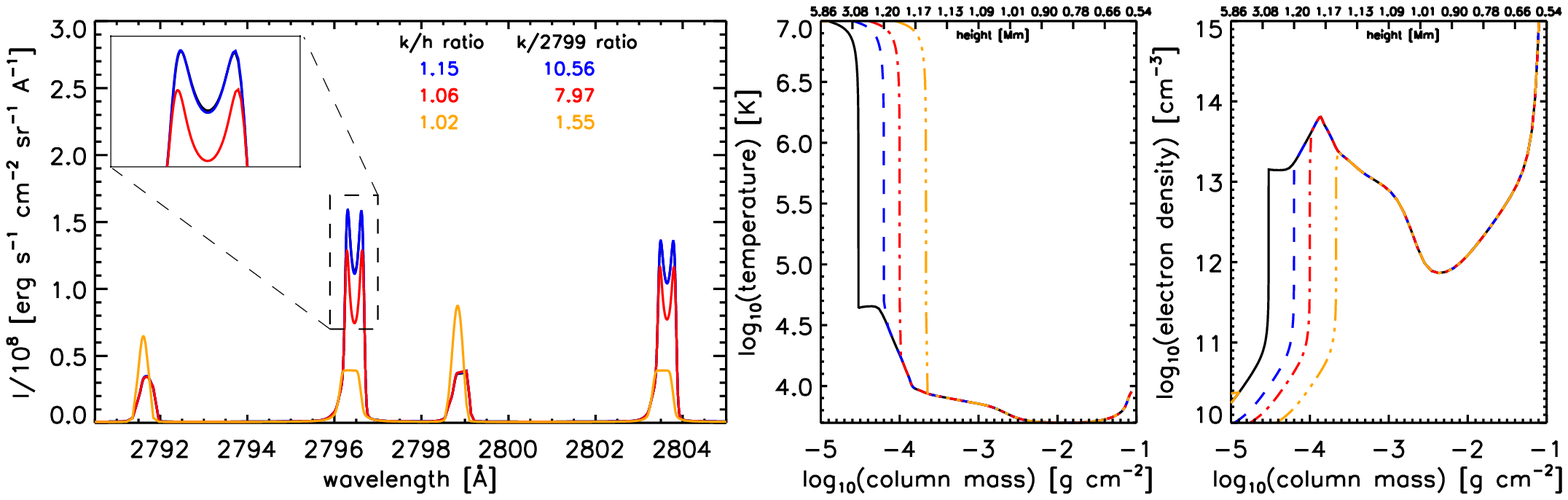}
   \caption{Variation of the transition region, simultaneously for the temperature and electron density. The non-thermal broadening and Doppler velocities were not changed. None of these variation lead to realistic flare profiles. Either the central reversal remains (blue, red), or in case of the yellow example, the subordinate lines are higher than the \ion{Mg}{2}~h~and~k lines, which is unrealistic.}
  \label{movetr}
\end{figure*}

\begin{figure*}[!tb]
\centering
  \includegraphics[width=\textwidth]{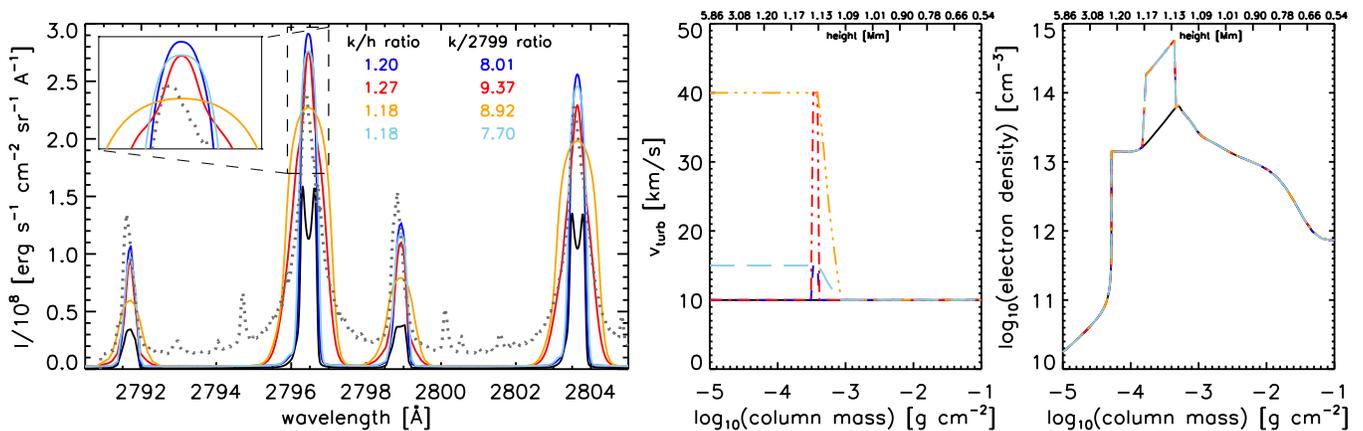}
   \caption{Increase of the non-thermal broadening in the upper chromosphere, up to (probably unrealistic) 40~km~s$^{-1}$. The electron density was taken from the red atmosphere of Figure~\ref{vard} to show the effect of the microturbulence parameter on an already existing single-peak profile. None of these examples show broad enough wings compared to the observations and additionally, high microturbulent values unrealistically broaden the line core.}
  \label{vturb}
\end{figure*}

To study how the emission of the \ion{Mg}{2}~k line responds to the combination of plasma motions in opposite directions, we combined a strong downflow (red line in Figure~\ref{combine2}) with a moderate upflow (blue) into a new profile (yellow, consisting of 35\% blue + 65\% red). The existence of unresolved up- and downflows in the same pixel could explain the broad wings. We increased the turbulent velocity to 15~km~s$^{-1}$, otherwise the absorption feature (as seen in Figure~\ref{combine}) affected the shape of the line wings. Also, the downflow velocities have to be larger than the upflows, because the red wing is slightly more extended in observations. Since nearly all {\it IRIS} flare ribbon pixels have these typical profiles, this result may indicate that there are similar unresolved up- and downflows throughout the ribbons.

   \subsection{Varying the radiation field}
By artificially including a UV radiation field in the upper corona in the RH code, including emission lines such as \ion{He}{1}, \ion{He}{2} and \ion{Mg}{10}, we aim at compensating for energy losses of optically thin lines not considered in the RH code, which may influence the resulting synthetic emission during solar flares, especially in the UV.

We considered a wavelength range between 302 and 613~\AA\ and estimated the total irradiance \citep{2004ApJ...606.1239P, 2004ApJ...606.1258J}. Comparing the resulting \ion{Mg}{2}~h and k line profiles with the original ones without external irradiation field in the upper corona, we found that the differences are minimal, especially in the wings. The emission in the line core is 0.02\% higher when including the external irradiation field in the upper corona. Therefore, the inclusion of coronal irradiance does not contribute enough to the total \ion{Mg}{2} UV emission, to explain the differences between the synthetic line profiles and those observed by {\it IRIS}.

   \subsection{Varying the transition region}
For this set of models we vary the location of the transition region by modifying the column mass at which the transition region is located. Figure~\ref{movetr} shows the modifications of temperature and density. The lower the transition region, the lower the intensity of \ion{Mg}{2}~h~and~k. These variations do not reproduce typical flare profiles. Even the yellow single-peaked profiles are not similar to observed profiles because the subordinate lines have higher intensities than the h and k lines.

   \subsection{Varying the non-thermal broadening}\label{Sect:vturb}
Since the resulting synthetic line profiles are narrower than the observed flaring \ion{Mg}{2} profiles, we also tried to modify the non-thermal broadening parameter, by increasing it in the upper chromosphere, at a column mass higher than -3.1~g~cm$^{-2}$. Figure~\ref{vturb} shows the broadening of the line profile for different test runs, while increasing the non-thermal broadening and varying the region where this occurs. For these simulations, we also increased the electron density in a narrow layer at the upper chromosphere, where the line core is formed, to obtain a single core line emission to be able to better compare the effects of microturbulence to observed flare profiles.

Considering that typical values for microturbulence to compensate for the lack of small-scale random motions in the model are $\sim 3$~km~s$^{-1}$ \citep{2012A&A...543A..34D, 2016ApJ...830L..30D}, even unrealistically high velocities of 40~km~s$^{-1}$ are not enough to explain the wide broad wings observed by {\it IRIS}. For instance \citet{2016ApJ...830L..30D} use 8~km~s$^{-1}$, describing this value as very high. Moreover, the line profile not only becomes broader, but also flatter in the core of the line, contrary to the observations.

\subsection{Varying the flux of non-thermal electrons}
We studied how the variation of the non-thermal electron flux influences the \ion{Mg}{2}~UV emission by using atmospheres from different RADYN simulations and calculating the profiles with RH. The resulting atmospheres correspond to four different simulations with a constant total non-thermal electron energy flux of $10^9$, $10^{10}$, $10^{11}$ and $10^{12}$ erg~cm$^{-2}$~s$^{-1}$ and assuming that the electron spectra follow a single power law at non-thermal energies, with a cutoff energy of $E_c=25$~keV and a spectral index $\delta=4$ for the first three simulations and $\delta=3$ for the simulation with the highest electron energy flux.

\begin{figure}[!tb]
\centering
 \epsscale{1.2}
  \plotone{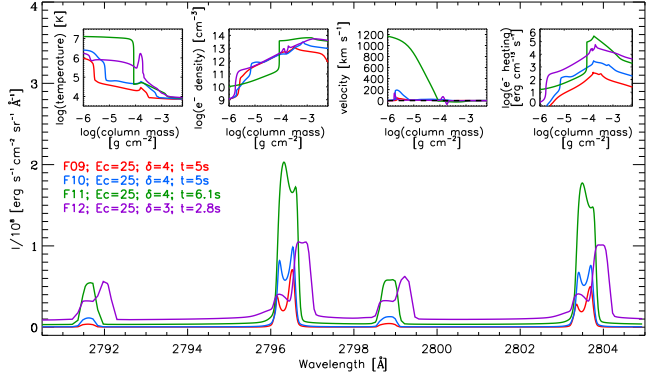}
  \vspace{-0.2cm}
   \caption{\ion{Mg}{2} spectra resulting from different atmospheres as a result of the RADYN simulations, varying the non-thermal electron flux. A very high flux (10$^{12}$ erg cm$^{-2}$~s$^{-1}$) leads to unusual profiles with a strong red component, however they do not match the observations. None of these simulations showed single-peak profiles at any time step.}
  \label{mgii_varF}
\end{figure}

Since the atmosphere evolves differently for each electron energy flux simulation, it is hard to make a direct comparison of the line profiles and assess a direct contribution of the electron flux. We tried to minimize this effect by selecting a timestep at the beginning of the simulation, when the atmosphere has not reached the ``explosive'' chromospheric evaporation described by \citet{2005ApJ...630..573A}. We selected a timestep of $t=5$~seconds after heating the atmosphere for the first two runs (F09 and F10), at which point the atmosphere has not reached the ``explosive'' chromospheric evaporation; $t=6.1$~seconds in the case of F11, having very strong upflow velocities in the lower corona of almost 1200 km~s$^{-1}$ as well as downflowing velocities close to 25 km~s$^{-1}$; and $t=2.8$~seconds for F12, with downflowing velocities of 70 km~s$^{-1}$. Figure~\ref{mgii_varF} shows how much the atmospheres differ within the runs.

In general, increasing the beam strength leads to higher intensities in the \ion{Mg}{2}~UV wavelength range, especially in the continuum regime as well as the subordinate line profiles. The asymmetries (higher red peaks) of the F09 and F10 runs (red and blue lines of Figure~\ref{mgii_varF}) are mostly due to the downflowing velocities (negative values) at a column mass region of $\approx -4$~g~cm$^{-2}$, where the line core is formed. The line profile resulting from the F11 run (green line) is the strongest among the four cases, having also the highest electron density and temperature, more than one order of magnitude increase in the upper chromosphere, between -3.4 and -4~g~cm$^{-2}$. The strong slightly blueshifted line core is due to the upflowing (positive) velocities of almost 25~km~s$^{-1}$ in the upper chromosphere. The F12 simulation (violet line) shows lower intensities in the line core and a lack of reversal at 2796.35~\AA\ due to the sudden temperature increase below a column mass of -4~g~cm$^{-2}$, decoupled from the electron density, since the sudden temperature increase does not reflect such sudden changes in the electron density. The secondary emission peak at higher wavelengths (2796.8~\AA) is due to the sudden changes in the velocity stratification, from upflowing to downflowing plasma motions, located at the column mass of -4~g~cm$^{-2}$. These changes are associated to the sudden temperature increase, located at that height.

Even though we only show the line profile from a specific time step of the RADYN simulations, we have performed a study of the temporal evolution of the line profiles. We found that none of the simulations showed \ion{Mg}{2}~h~and~k as single-peaked profiles for any beam strength at any time step.

   \subsection{Including particle transport and acceleration}
By considering the coupling of particle transport and stochastic acceleration using the FLARE code \citep{2002ApJ...569..459P} with the radiative transfer and hydrodynamics from the RADYN code, we obtain a spectrum formed by a quasi-thermal component and a power-law tail as explained by \citet{2015ApJ...813..133R}. The atmospheric evolution of the run when coupling of the particle transport and acceleration equations with the hydrodynamics results in a stronger chromospheric evaporation as well as greater up- and downflowing velocities, when compared with the atmosphere resulting from the single power-law run.

\begin{figure}[!tb]
\centering
 \epsscale{1.25}
  \plotone{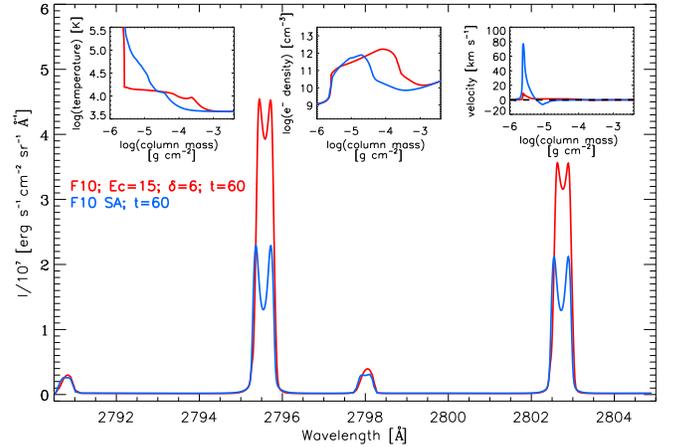}
  \vspace{-0.2cm}
   \caption{\ion{Mg}{2} spectra resulting from the PL and SA1 runs of \citet{2015ApJ...813..133R}, at the peak of the electron energy flux, after 60 seconds of heating. SA leads to lower intensities, but single-peaked profiles cannot be reproduced with either method.}
  \label{mgii_var_SA_PL}
\end{figure}

By taking the atmosphere from the hydrodynamic simulations of \citet{2015ApJ...813..133R} after 60 seconds of heating, we studied how the \ion{Mg}{2}~h~and~k~line profiles are affected when coupling particle transport and stochastic acceleration. We find that although the intensity decreases by almost a factor of 2, the line profiles still show narrow wings and a core reversal at the line center (see~Figure~\ref{mgii_var_SA_PL}).

\begin{figure*}[!htb]
\centering
  \includegraphics[width=.49\textwidth]{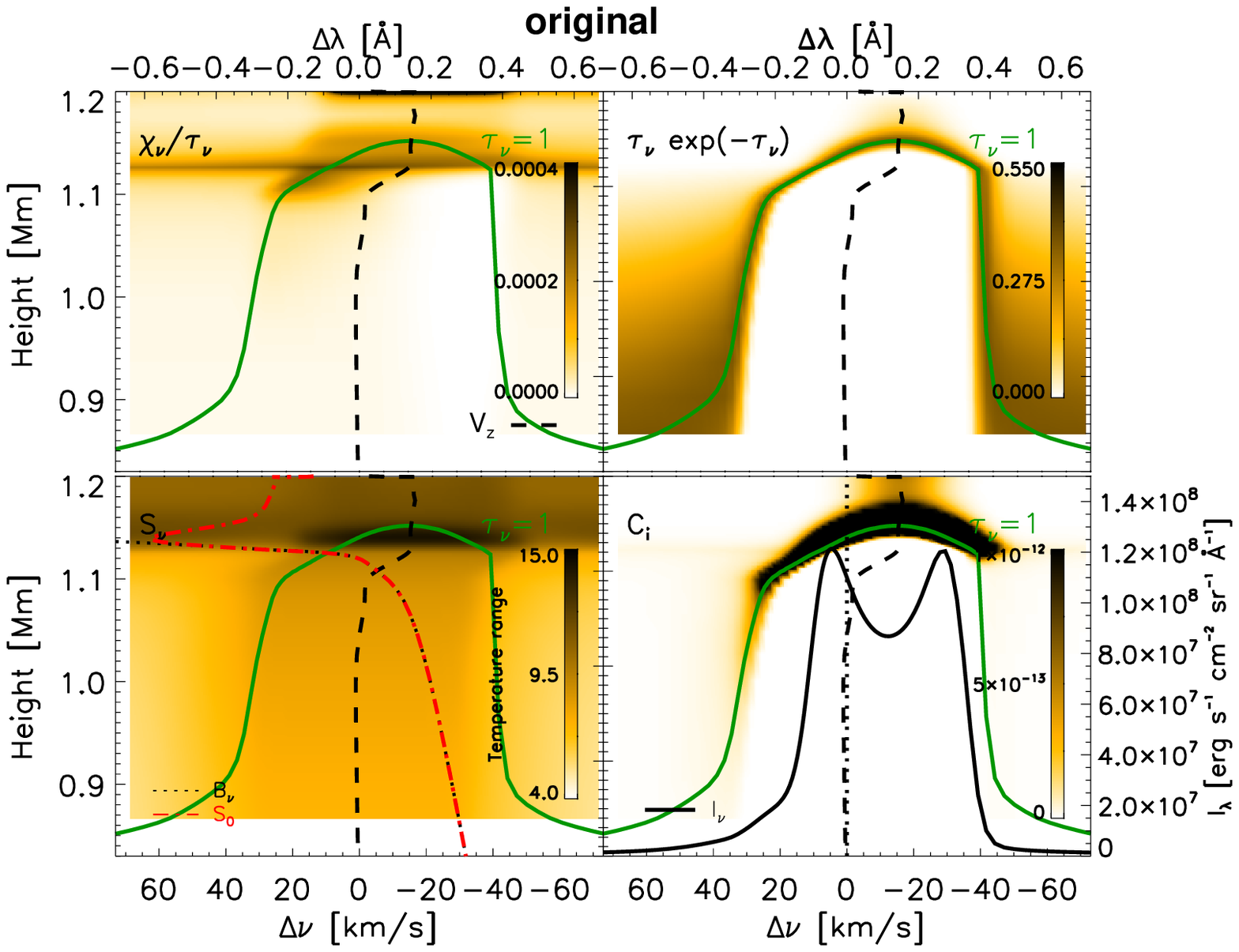}
  \includegraphics[width=.49\textwidth]{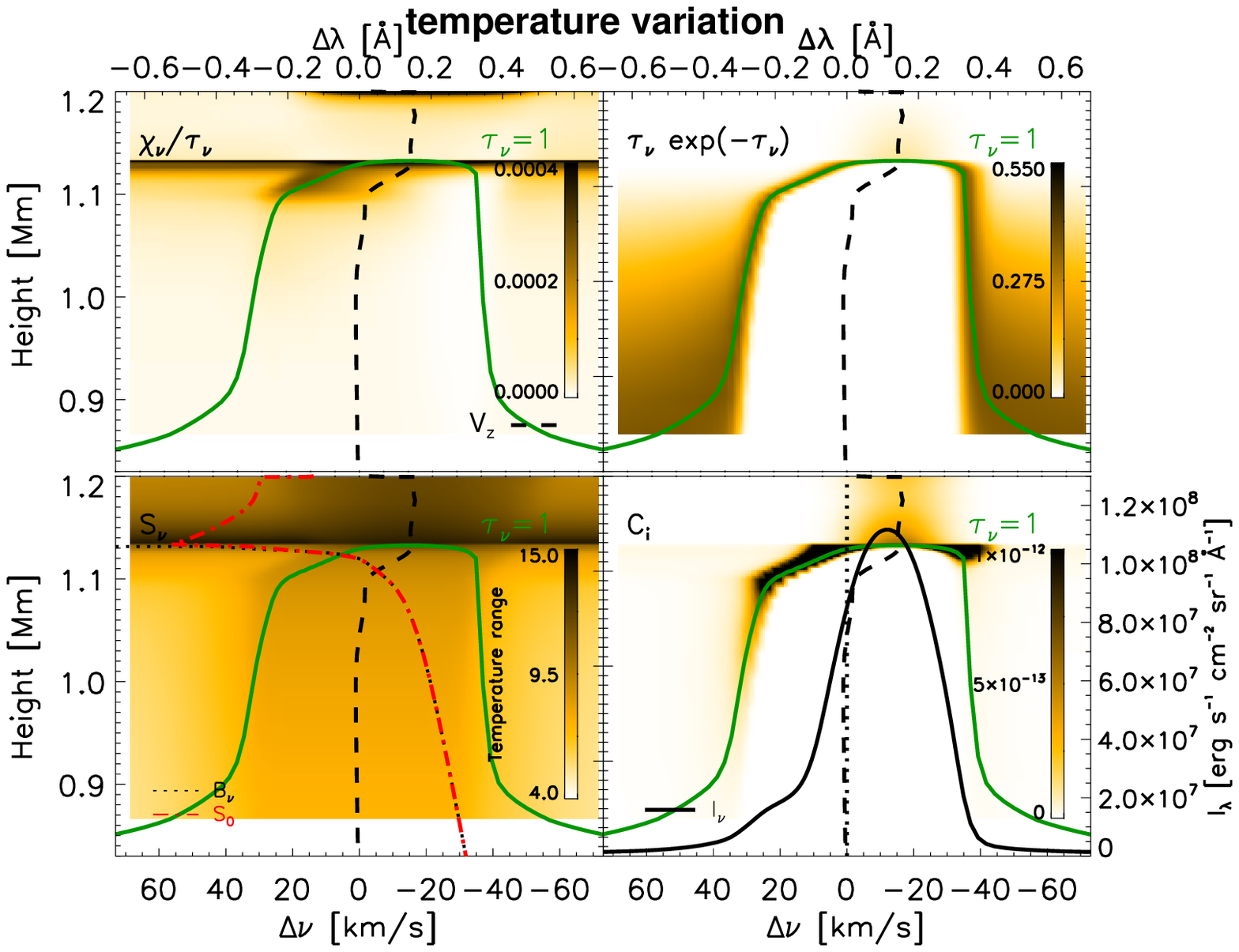}
  \includegraphics[width=.49\textwidth]{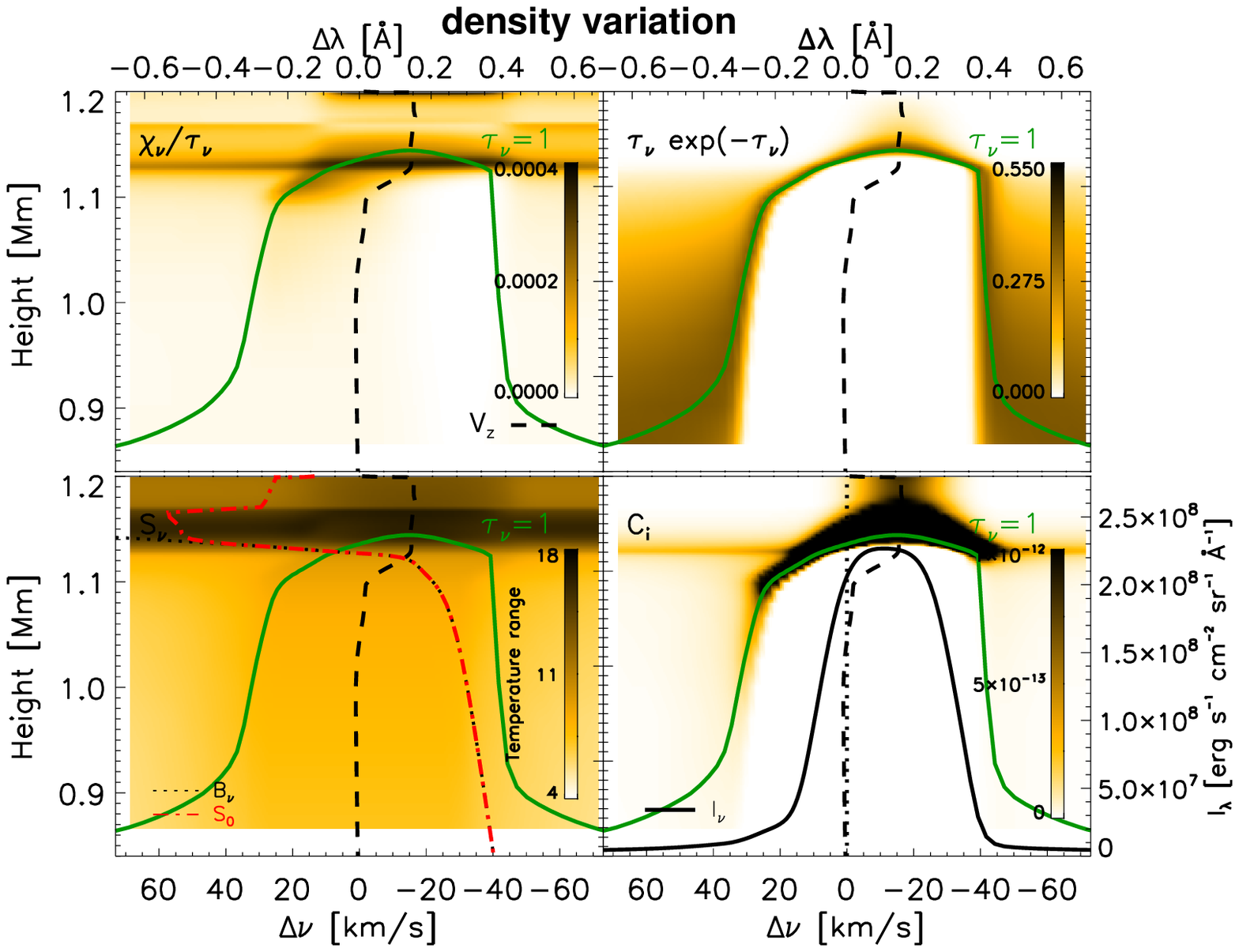}
  \includegraphics[width=.49\textwidth]{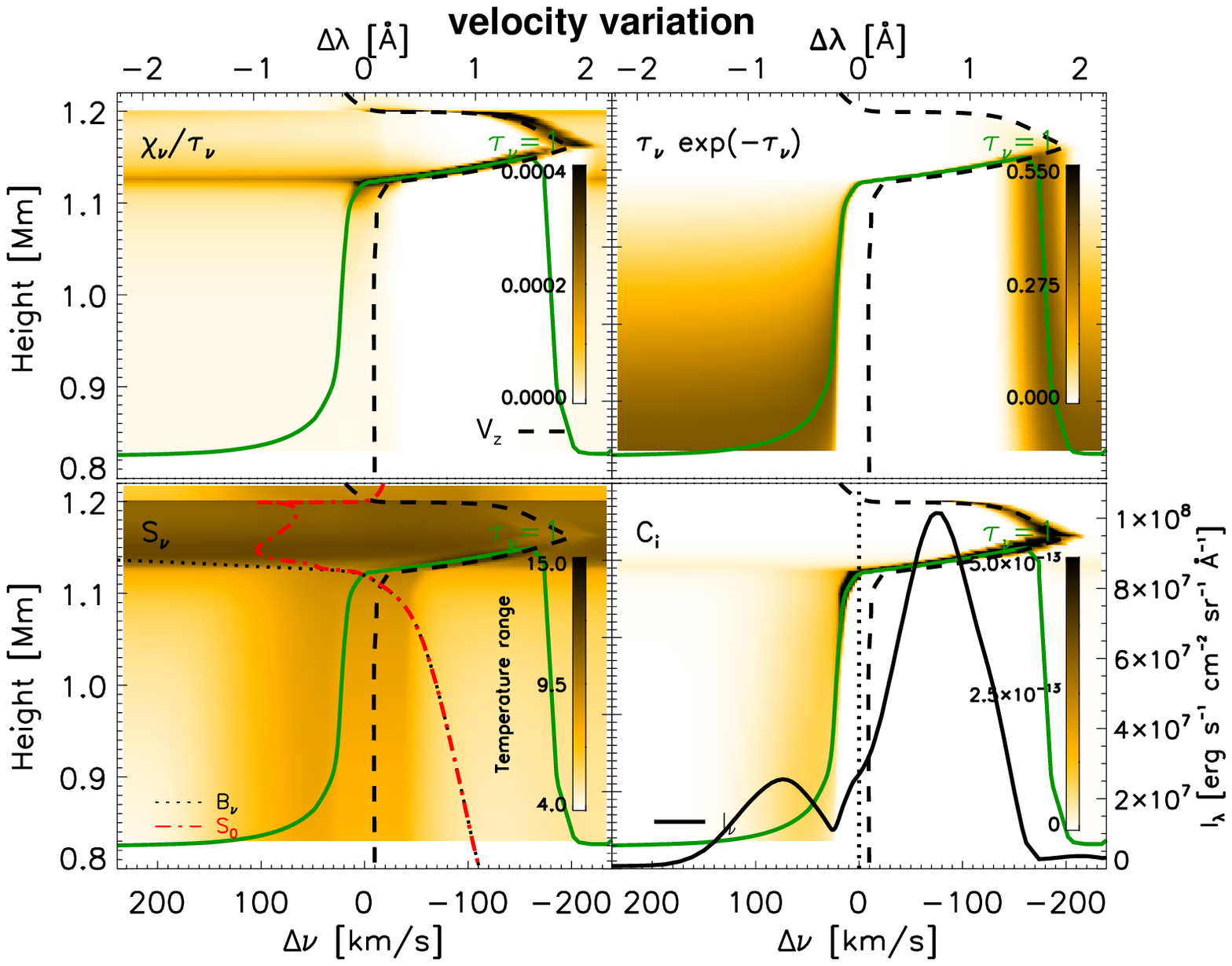}
   \caption{Intensity contribution function resulting from the different atmospheres. Darker regions indicate areas of strong contribution. The green line shows the $\tau_{\nu}=1$ location and the black dashed line shows the vertical velocity as a function of height. The black dotted line in the lower left panel represents the Planck function $B_{\nu}$ and the red dashed-dotted line, the source function $S_{\nu}$ calculated at the line center. The black line in the lower right panel show the spectral line profile. Top left: $C_i$ from the original RADYN atmosphere, without any modification. Top right: $C_i$ from the modified RADYN atmosphere, after increasing the temperature (yellow atmosphere from Figure~\ref{vart}). Bottom left: $C_i$ from the modified RADYN atmosphere, after increasing the density (yellow atmosphere from Figure~\ref{vard}). Bottom right: $C_i$ from the modified RADYN atmosphere, after increasing the plasma velocity (dark blue atmosphere from Figure~\ref{combine}).}
  \label{ci_images}
\end{figure*}

The line profiles resulting from the power-law run (red line) have stronger intensities in the line core, but almost the same contribution in the continuum as the stochastic acceleration run. This is because, although the temperature and electron densities are about the same order of magnitude, the atmosphere resulting from the power-law run shows slightly higher densities and temperatures at a column mass between -3.4 and -4~g~cm$^{-2}$.

The inclusion of particle transport and acceleration results in higher upflowing plasmas close to the transition region (below to a column mass of -6~g~cm$^{-2}$) and stronger downflowing velocities in the upper chromosphere (at a column mass of -5~g~cm$^{-2}$). The line core is formed lower in the atmosphere, at a column mass close to -3.6~g~cm$^{-2}$; therefore the plasma motions that significantly differ in both runs do not affect the line formation. Instead, the line profiles are symmetric, as the synthetic spectra in Figure~\ref{mgii_var_SA_PL} show.

  \section{Formation of the \ion{Mg}{2}~k line profile: intensity contribution function}\label{Sect:contr_fct}
To investigate the behavior of the source function and where in the atmosphere the line is formed, we calculate the so-called contribution function. By writing the formal solution of the transfer equation for emergent intensity \citep[Equation~\ref{eq_contribution_function}; ][]{1997ApJ...481..500C}, we can investigate how the atmospheric stratification affects the formation of the line profile and where exactly the line is formed. The line emergent intensity $I_{\nu}^{0}$ can be written as:

\begin{equation}
I_{\nu}^{0} =
\frac{1}{\mu} \int_{z} S_{\nu} \chi_{\nu} e^{-\frac{\tau_{\nu}}{\mu}} dz = \frac{1}{\mu} \int_{z} C_i \,dz \,\,,
\label{eq_contribution_function}
\end{equation}
where $z$ is the atmospheric height; $\tau_{\nu}$ is the monochromatic optical depth; $\chi_{\nu}$ is the monochromatic opacity per unit volume; $S_{\nu}$ is the source function, defined as the ratio between the emissivity to the opacity of the atmosphere and $C_i$ is the intensity contribution function, indicating how much emergent intensity originates at a certain height $z$.

The intensity contribution function, $C_i$, can be expressed as the product of $S_{\nu}$, $\frac{\chi_{\nu}}{\tau_{\nu}}$ and $\tau_{\nu}e^{-\frac{\tau_{\nu}}{\mu}}$, as the panels in Figure~\ref{ci_images} show \citep[for a more detailed explanation see][]{2013ApJ...772...90L}. Since both \ion{Mg}{2}~h and k line profiles have a similar behavior, we will focus on the \ion{Mg}{2}~k line profile in the following sections.

  \subsection{Original RADYN atmosphere}\label{Sect:contr_fct_orig}
The \ion{Mg}{2}~k line profile resulting from the simulation using the original RADYN atmosphere shows a reversal at the line center (top left panel of Figure~\ref{ci_images}). The asymmetry in the blue wing of the line is associated to the asymmetry in the $\tau_{\nu}=1$ layer at different frequencies, as the term $\tau_{\nu}e^{-\frac{\tau_{\nu}}{\mu}}$ in the upper right panel indicates, since its contribution to the total intensity at each frequency has an asymmetry towards blue wavelengths. The source function $S_{\nu}$ in the lower left panel, calculated for the line center wavelength, closely follows the Planck function (dotted line) in the lower atmosphere, up to 1.14~Mm in comparison with the sudden increase at higher layers, due to the temperature stratification. The departure of the source function from the Planck function means that the line core is formed under non-LTE conditions, because both profiles have a different temperature response.
   
The intensity contribution function $C_i$ is the product of the previous three panels: the main contribution comes from the upper chromosphere, between 1.06 and 1.14~Mm, since the ratio of the opacity to the optical depth ($\frac{\chi_{\nu}}{\tau_{\nu}}$) is very small at lower heights. The integration of the intensity contribution function at each frequency results in the black line profile displayed in the lower right panel. The core reversal is due to the sudden decrease of the source function due to the decoupling of the source and Planck functions at 1.14~Mm, where the photons of the line core are formed.

By studying the formation of the subordinate lines between the 3p$^2P$ and 3d$^2D$ states (figure not shown), we found that they are formed in the upper chromosphere, between 0.83 and 1.14~Mm, at higher layers than the values reported by \citet{2015ApJ...806...14P} for the quiet Sun.

  \subsection{Atmosphere with increased temperature}\label{sectempatmos}
By increasing the temperature in the upper chromosphere, as described by the yellow atmosphere of Figure~\ref{vart}, we find that the formation height of the line center is reduced by $\sim$200~km (see top right panel of Figure~\ref{ci_images}), explaining the steeper peak at the line core. The $\tau_{\nu}=1$ layer in the line wings has the same shape as in the top left panel of Figure~\ref{ci_images} with the exception of the plateau at wavelengths close to the line core. The redshifted line core is due to the downflowing velocities of $\sim$~14~km~s$^{-1}$ present at the line core formation height (between 1.06 and 1.14~Mm). Although the ratio between the opacity and the optical depth is almost the same as in the previous atmospheric example, there is no clear emissivity above $\tau_{\nu}=1$ close to the line center. The temperature increase results in a coupling of the source function to the Planck function at higher heights, especially also where the line core is formed (lower left panel). This indicates LTE conditions for the whole line and explains the single-peaked profile.

The asymmetry of the line profile is reflected by the frequency distribution of the intensity contribution function. There is a secondary intensity contribution function component at a height of 1.13~Mm, and above $\tau_{\nu}=1$, which indicates that there is some contribution to the core emission, which is formed under optically thin conditions.

  \subsection{Atmosphere with increased density} \label{Sect:contr_fct_dens}
The bottom left panel of Figure~\ref{ci_images} shows the different terms of the total intensity contribution function for the yellow atmosphere of Figure~\ref{vard}. Here the electron density increase results in a slightly stronger opacity component $\chi_{\nu}$ at a height of 1.14~Mm, in comparison with the resulting opacity of Section~\ref{sectempatmos}, with an increased atmospheric temperature. The term $\chi_{\nu}/\tau_{\nu}$ is also stronger in the line center, in comparison with the original atmosphere of Section~\ref{Sect:contr_fct_orig} because $\chi_{\nu}$ is proportional to the density of emitting particles. Therefore, $\frac{\chi_{\nu}}{\tau_{\nu}}$ is higher when there is a large number of emitters at low optical depth (i.e., the produced photons can escape), explaining the stronger emissivity component in comparison with the top right panel of Figure~\ref{ci_images}.

The main difference in comparison with the increase of temperature is that the coupling between the source function and the Planck function is located at slightly higher heights, as well as the formation height ($\tau_{\nu}=1$) of the line core. By increasing the electron density, the source function $S_{\nu}$ keeps increasing with the temperature, although it is decoupled from the Planck function $B_{\nu}$. As in Section~\ref{sectempatmos}, the source function is still coupled to the Planck function at the core formation height, indicating LTE conditions. The line core shows a flatter profile because the intensity contribution function has a more symmetric frequency distribution, with a fainter blue wing asymmetry. 

  \subsection{Atmosphere with increased velocities} \label{Sect:contr_fct_vel}
Previous simulations have shown that asymmetries in chromospheric line profiles during flares are strongly affected by the velocity field in the flaring atmosphere \citep{2015ApJ...813..125K}. The bottom right panel of Figure~\ref{ci_images} shows that by increasing the downflowing plasma velocity to about 200~km~s$^{-1}$, as shown in the blue atmosphere of Figure~\ref{combine}, the line profile is formed in a broader region, in comparison with the original atmosphere of the top left panel of Figure~\ref{ci_images}. The intensity contribution function also covers a much broader frequency range, broadening the wings of the line profile (note the different scaling of the x-axis). The Planck and source functions have similar behavior, increasing with temperature in the lower atmosphere, departing from the Planck function above 1.12~Mm, similarly to the original atmosphere. At the core formation height, just above 1.12~Mm, the source function is starting to decouple from the Planck function, although both still increase with temperature, resulting in a single core in emission. Due to the plasma motions, the line shows an asymmetric profile, especially towards redder wavelengths, as a result of the increased downflowing plasma motions.

By increasing the plasma velocity, the formation height at which $\tau_{\nu}=1$ (green line) is not flat near the line core, as in the previous cases, but it linearly increases within the wavelength range $-0.2$ to $1.6$~\AA. 

For other variations of the velocity, we can either have still coupled Planck and source functions (e.g. red line in Figure~\ref{varv}), or an already decoupled and decreasing source function with temperature (e.g. red line in Figure~\ref{combine}). Yet we do not observe a reversal in any of the cases because even with a decreasing source function, the large velocities fill the intensity where usually a reversal is observed.

It is also worth noting the secondary contribution coming from heights close to 1.2~Mm (not shown), above the $\tau_{\nu}=1$ curve, indicating that there is a contribution to the intensity in the red wing (at wavelengths between 0.6 and 2.1~\AA) formed under optically-thin conditions.

  \section{Summary and Discussion}\label{sect:conclusions}
   \subsection{Options to simulate flare-like Mg profiles}
Our parameter study has shown several methods to obtain more flare-like \ion{Mg}{2}~h~and~k profiles in simulations with a single peak in emission. In Section~\ref{Sect:vart} we have seen that by increasing the temperature at the line core formation height, we do not only obtain a single peak in emission, but also decrease the peak intensity. This is favorable, considering that synthetic line profiles tend to show higher intensities than {\it IRIS} observations \citep{2016ApJ...827...38R, 2017ApJ...836...12K}. This modification also results in symmetric line profiles, similar to the typical observed flare ribbon profiles. While the line ratio between the \ion{Mg}{2}~k and \ion{Mg}{2}~h lines looks reasonable for most of our simulations, the ratio between them and the subordinate triplet lines is too large for the atmosphere with a modified temperature (cf. Figure~\ref{vart}). This may indicate that the temperature or its gradient in the lower chromosphere may also need to be modified. However, this can be tricky as it might affect other lines, such as H\textalpha\ or \ion{Ca}{2}~8542~\AA, which already show a good agreement with observations (see Figure~12 in \citet{2015ApJ...804...56R}, or Figure~11 in \citet{2016ApJ...827...38R}, for an easier comparison, we marked the locations in Figure~\ref{examplemg}).

The increase of the electron density results in a line core in emission as well, but as can be seen in Figure~\ref{vard}, the intensity of the line increases significantly, to values not reached by observations. The line ratio of \ion{Mg}{2}~k to subordinate lines is closer to the observations for the electron density variation than for the temperature variation.

In the case of increased downflowing velocities, the line core may show a single-peak in emission as well. The formation height at which the optical depth $\tau_{\nu}=1$ increases linearly near the center of the line, with its peak at 1.15~Mm and 1.6~\AA. At this height, $S_{\nu}$ and $B_{\nu}$ are already decoupled. The source function at the line core formation heights for the velocity variations can either be coupled to or decoupled from the Planck function, depending on the velocity stratification.

The intensity contribution function $C_i$ shows a secondary component in emission, formed at heights above $\tau_{\nu}=1$ and therefore under optically-thin conditions. However, the resulting line profile is highly asymmetric, contrary to standard flare profiles. Only by combining profiles with different velocity structures, one can reproduce different {\it IRIS} profiles, thus indicating that the {\it IRIS} profiles may be unresolved. By combining downflows velocities, starting at different heights, the shape of the short-lived {\it IRIS} profile from HXR emission sites can be reproduced. The single peak and broad wings of typical {\it IRIS} flare profiles can be reproduced by combining up- and downflows. The unresolved components could also be explained by a combination of several threads overlaying in space \citep{2006ApJ...637..522W, 2016ApJ...827...38R, 2016ApJ...827..145R}. The subordinate lines are comparably low in this case, and would require further modifications in the lower chromosphere as discussed above.

\begin{figure}[!tb]
\centering
  \includegraphics[width=.45\textwidth]{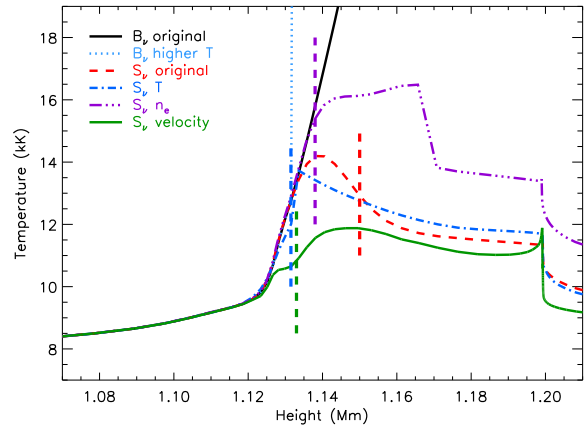}
  \vspace{-0.4cm}
   \caption{Planck and source functions resulting from the three different atmospheres discussed in Section~\ref{Sect:contr_fct}. The vertical dashed line marks the formation height of the \ion{Mg}{2}~k line core for each atmosphere.}
  \label{src_fct}
\end{figure}

By increasing the temperature or density (Section~\ref{Sect:vart} and \ref{Sect:vard} and purple and blue lines in Figure~\ref{src_fct}), the line core is formed under LTE conditions; while the velocity variations may result in a decoupling of the source function from the Planck function, therefore non-LTE conditions (see green line in Figure~\ref{src_fct}). In these cases we still have single peak profiles due to the large asymmetries of the line. It is possible that these conditions occur during flares, possibly through a combination of all parameters discussed above.

  \subsection{Options that fail to simulate flare-like Mg profiles}
The increase of the chromospheric temperature (Figure~\ref{truns}) as well as the shift of the transition region towards lower heights (Figure~\ref{movetr}) does not result in realistic line profiles. Either the line core is still reversed, or the subordinate lines have stronger intensities than those of the \ion{Mg}{2}~h~and~k lines, showing a discrepancy with respect to {\it IRIS} flare observations.

As discussed in Section~\ref{Sect:vturb}, we also discarded a higher non-thermal broadening as a possibility to explain the broad observed line wings. Although the Van der Waals broadening seems to be the main contribution to the broad line wings, this term does not explain the widely broadened observed line wings. Although the RH code considers quadratic Stark broadening in the calculation, the adiabatic approximation used in RH for the quadratic Stark effect for the Mg atom underestimates the broadening \citep{1978stat.book.....M}. Therefore, we should keep in mind that the implementation of the quadratic Stark effect may not be sufficiently accurate. Very wide \ion{Mg}{2} line profiles have also been observed in stellar flares \citep[i.e. on YZ CMi, observed by][]{2007PASP..119...67H}. \citet{2007PASP..119...67H} discussed non-thermal beam excitations as the reason for the broadening of the \ion{Mg}{2} wings. They argue that unresolved flows with high velocities ($\sim$~200~km~s$^{-1}$), which would result in high kinetic energies that may exceed the radiated energy in the flare, are a possible explanation of the broad wings. Further studies with higher non-thermal electron fluxes will clarify this point.

A larger opacity in the photosphere would result in broader wings and this could be accomplished by increasing the temperature or the density, but would probably affect other spectral lines whose fit currently is reasonably good. 3D effects, as discussed by \citet{2013ApJ...772...89L}, may have an important contribution on the formation of the \ion{Mg}{2}~h~and~k line emission, especially in the line core, but are currently not computationally feasible.

The importance of the advection term in the statistical equilibrium equation for \ion{Mg}{2} it is still unclear. Cool flows flowing into hot plasma or hot flows into cool plasma could result in non-equilibrium of the ionization and excitation.

We would like to emphasize that this is purely a parameter study and we do not claim that the density or temperature must increase by an order of magnitude in flares. We do not have a physical explanation yet as to what is occurring in a flaring atmosphere. Nevertheless, we can conclude that a critical piece is missing in current hydrodynamic simulations, possibly more particles that may increase densities, more heat dissipation, or unresolved and stronger velocities. For future work, we plan to simultaneously compare many spectral lines, which will better constrain the atmospheric parameters at different heights and will potentially allow us to better constrain the conditions in flaring atmospheres.

\acknowledgments
We would like to thank P. Judge, P. Heinzel, W. Liu, R. Rutten, G. Kerr, M. Carlsson, T. Pereira, and the anonymous referee for their helpful discussions. Work performed by F.R.dC. is supported by NASA grants NNX13AF79G, NNX14AG03G, 8100003073 and NNX17AC99G. L.K. was partially supported by NASA grant NNX13AI63G. F.R.dC. and L.K. thank ISSI and ISSI-BJ for the support of the team ``Diagnosing heating mechanisms in solar flares through spectroscopic observations of flare ribbons''. We gratefully acknowledge the use of supercomputer resources provided by the NASA High-End Computing (HEC) Program through the NASA Advanced Supercomputing (NAS) Division at Ames Research Center.
{\it IRIS} is a NASA small explorer mission developed and operated by LMSAL with mission operations executed at NASA Ames Research center and major contributions to downlink communications funded by ESA and the Norwegian Space Centre. CHIANTI is a collaborative project involving George Mason University, the University of Michigan (USA), and the University of Cambridge (UK). 

\bibliographystyle{apj}
\bibliography{ads}
\end{document}